\documentclass[aps,prd,reprint]{revtex4-2}
\usepackage{graphicx}
\usepackage{times}
\usepackage{bm}
\usepackage{amsmath,amssymb}
\usepackage{feynmf}
\usepackage{multirow}

\begin{document}
\title{Finite time effects in single and double Compton scattering}
\author{V. Dubrovich$^{1}$, T. Zalialiutdinov$^{2, 3}$,}
\affiliation{
$^1$ Special Astrophysical Observatory, St. Petersburg Branch, Russian Academy of Sciences, 196140, St. Petersburg, Russia\\
$^2$ Department of Physics, St. Petersburg State University, Petrodvorets, Oulianovskaya 1, 198504, St. Petersburg, Russia\\
$^3$ Petersburg Nuclear Physics Institute named by B.P. Konstantinov of National Research Centre "Kurchatov Institut", St. Petersburg, Gatchina 188300, Russia
}
\email[E-mail:]{t.zalialiutdinov@spbu.ru}

\begin{abstract}
The process of Compton scattering by a free electron with subsequent reemission of one or two photons is considered in the assumption of finite interaction time. The corresponding cross sections are obtained in the framework of relativistic quantum electrodynamics using a modified form of fermion propagator with complex transmitted momentum. It is shown that finite time effects can be observable at sufficiently low energies of scattered photons. 
The proposed method also regularizes arising infrared divergence in the cross section of the double Compton effect. Possible experimental verification of considered theoretical approach is discussed.
\end{abstract}

\maketitle

\section{Introduction}

In this paper, we consider the derivation of single and double Compton scattering cross section, assuming that the incoming photon interacts with the electron for only a finite time $ \tau$. The explicit introduction of time-dependent adiabatic factors at the interaction vertex immediately makes the theory no longer Lorentz-invariant \cite{Sucher}. It complicates the construction of quantities such as the transition probabilities and the scattering cross sections. However, the approach developed for interactions with unstable particles proves helpful here. Recently, it was shown that for unstable fermion particles, the following modified Feynman propagator might be introduced \cite{CASTRO,CASTRO2,Nowakowski1993,Kuksa2015}
\begin{eqnarray}
\label{1}
S_{F}(p)=\frac{\hat{p}+m_{F}-\mathrm{i}\Gamma/2}{p^2-(m_{F}-\mathrm{i}\Gamma/2)^2}
,
\end{eqnarray}
where $ p $ is the four-vector of fermion momentum, $ m_{F} $ is the mass of fermion, $ \hat{p} = \gamma^{\mu}  p_{\mu} $, $ \Gamma $ corresponds to the decay width of the particle and, in general, could depend on transmitted momentum $q$. In contrast to the naive introduction of factor $\Gamma$ in the numerator of ordinary Feynman propagator \cite{weinberg1995quantum}
\begin{eqnarray}
\label{2}
S_{F}(p)=\frac{\hat{p}+m_{F}}{p^2-m_{F}^2-\mathrm{i}m_{F}\Gamma}
,
\end{eqnarray}
the representation Eq. (\ref{1}) satisfy the electromagnetic Ward identity, which provides the gauge-invariant description of processes \cite{Kuksa2014, Kuksa2015}. 

Parameter $\Gamma$ introduced in Eq. (\ref{1}) and associated with the particle decay implies the finiteness of time interaction in the scattering processes. The corresponding interaction interval is defined by the fermion decay time $ \tau=1/\Gamma $. In principle, the localaized in time effects for scattering of stable particles can be introduced in a similar manner. Then the corresponding  $\Gamma$ in Eq. (\ref{1}) can be associated with the duration time of laser pulse $ \tau_{L} $. This picture can be true for a single quasi-monochromatic photon with incident energies much higher than the width of photon non-monochromaticity. Without loss of generality, the imaginary addition to the momentum transferred from a photon to a fermion can be attributed either to the lifetime of an unstable particle, or to the interaction time (inverse width) of a quasi-monochromatic photon with a stable fermion.

It should be noted that the exactly similar situation arises in atomic physics for the scattering of photons by an atomic electron. In this case, parameter $\Gamma$ is associated with the natural width of atomic level, which can be rigorously introduced within bound state QED theory of line profile or within the quantum mechanical description of quasi-stationary states \cite{Andr, ZSLP-report}. However, in contrast to QED in external fields the small imaginary addition to the energy of free stable fermion can be introduced only phenomenologically. With such restrictions and with the use of propagator in the form of Eq. (\ref{1}), we try to construct the cross section for the scattering of quasi-monochromatic photon by a free fermion. 

The paper is organised as follows. In section \ref{single} fully relativistic QED derivation of single Compton scattering of quasi-monochromatic photon is presented. The section \ref{double} is devoted to the similar consideration of double Compton effect. The results and discussions are presented in section \ref{conclusion}.

Throughout this paper relativistic units are employed, where the velocity of light $ c=1 $, Planck's constant $ \hbar=1 $, respectively. The electron mass $ m_{e} $ is written explicitly. The electron charge $ e=-\left|e\right| $ is related to the fine structure constant $ \alpha $ via
 $\alpha = e^2/4\pi$. For $4-$vectors and tensor a standard notations are used for covariant (lower index) and contravariant (upper index) components, which are related to each other by metric tensor with Minkowski metrics $g_{\mu\nu}=g^{\mu\nu}=(1,-1,-1,-1)$, so that, for example, for vector components $a_{\mu}$: $a_{\mu}=g_{\mu\nu}a^{\nu}$ (implying here and further Einstein's sum convention). Greek indices run over set $(0,\,1,\,2,\,3)$ and Latin indices run over set $(1,\,2,\,3)$. The contraction of Dirac $ \gamma^{\mu} $-matrices ($\mu=0,\,1,\,2,\,3$) with four-vector $ a=(a_{0},\textbf{a}) $ is denoted as $ \hat{a}=\gamma^{\mu}a_{\mu}$.

\section{Single compton scattering}
\label{single}

We start from the standard QED derivation of Klein-Nishina formula \cite{Klein1929} for Compton process but with the modified propagator in the form given by Eq. (\ref{1}). The corresponding S-matrix element for the single photon scattering $ \gamma(k_1)+e^{-}\rightarrow \gamma(k_2) + e^{-} $ is \cite{bjorken1964relativistic, JauchRohrlich}
\begin{eqnarray}
\label{3}
S^{(2)}_{fi}=(2\pi)^4\delta^{(4)}(p_{f}+k_2-p_{i}-k_1)
\\\nonumber
\times
\frac{m_{e}e^2}{\sqrt{8 V^4 E_{i}E_{f}\omega _{1}\omega _{2}}}{\cal M}(k_1,k_2)
,
\end{eqnarray}
where $ p_{i}=(E_{i},\textbf{p}_{i}) $, $ p_{f}=(E_{f},\textbf{p}_{f}) $  are the four-vectors of initial and final electron momentum, respectively, $ k_1=(\omega_1,\textbf{k}_1) $ is the four vector of incident photon with frequency $\omega_1$ and wave-vector $ \textbf{k}_1 $, $ k_{2}=(\omega_{2},\textbf{k}_{2}) $ is the four-vectors of outgoing photon, and  $ \cal{M} $ is the Feynman amplitude of the process  
\begin{eqnarray}
\label{4}
{\cal M}(k_1,k_2)=
\overline{u}(p_{f})
\left\lbrace
\hat{\varepsilon}(k_2)
S_{F}(p_{i}+k_1)
\hat{\varepsilon}(k_1)
\right.
\\\nonumber
+
\left.
\hat{\varepsilon}(k_1)
S_{F}(p_{i}-k_2)
\hat{\varepsilon}(k_2)
\right\rbrace
u(p_{i})
.
\end{eqnarray}
In Eq. (\ref{4}) $u(p)$ is the Dirac spinor for free electron with its Dirac adjoint defined as $ \overline{u}(p)=u^{\dagger}(p)\gamma_{0} $, $ \gamma^{\mu} $ are the Dirac matrices, $\varepsilon(k)$ is the polarization four-vector of a photon with momentum $ k $ and $ S_{F} $ is the Feynman propagator given by the representation Eq. (\ref{1}). 

The transition rate per unit time to one define state can be found according to the definition \cite{bjorken1964relativistic}
\begin{eqnarray}
\label{5}
w=\frac{|S^{(2)}_{fi}|^2}{T}=V(2\pi)^4\delta^{(4)}(p_{f}+k_2-p_{i}-k_1)
\\\nonumber
\times
\frac{m_{e}^2e^4}{4 V^4 E_{i}E_{f}\omega _{1}\omega _{2}}|{\cal M}(k_1,k_2)|^2
,
\end{eqnarray}
where $ T\rightarrow \infty $ is the observation time and $ V $ is the phase volume. Since we are interested to a transition rate $ dw $ to a group of final states with momenta in the intervals $ (\textbf{p}_f,\textbf{p}_f + d\textbf{p}_f) $ and $ (\textbf{k}_2,\textbf{k}_2 + d\textbf{k}_2) $ we must multiply Eq. (\ref{5}) by the number of these states which is 
\begin{eqnarray}
\label{6}
\frac{V^2d^3\textbf{p}_{f} d^3\textbf{k}_2}{(2\pi)^6}
.
\end{eqnarray}

With our choice of normalization for the states, the volume $ V $ contains one scattering center and the incident photon flux is  $ F=c/V $ ($c$ is the speed of light) \cite{rapoport}. Then the corresponding differential cross section can be found as follows
\begin{eqnarray}
\label{7}
d\sigma_{\mathrm{sc}}=\frac{dw}{F}
=\delta^{(4)}(p_{f}+k_2-p_{i}-k_1)
\\\nonumber
\times
\frac{m_{e}^2e^4}{4 E_{i}E_{f}\omega _{1}\omega _{2}}|{\cal M}(k_1,k_2)|^2
\frac{d^3\textbf{p}_{f} d^3\textbf{k}_2}{(2\pi)^2}
.
\end{eqnarray}
Integration over $ d\textbf{p}_{f} $ in Eq. (\ref{7}) can be performed with the use of equality \cite{bjorken1964relativistic, Akhiezer}
\begin{eqnarray}
\label{8}
\frac{d^3\textbf{p}_f}{2E_f}=\int d^4p_f \delta(p_f^2-m_e^2)\theta(p_{f}^0)
,
\end{eqnarray}
where $ \theta $ is the Heaviside step-function. After that Eq. (\ref{7}) takes the form
\begin{eqnarray}
\label{9}
d\sigma_{\mathrm{sc}}
=\delta((p_i+k_1-k_2)^2-m_e^2)\theta(m_{e}+\omega_1-\omega_2)
\\\nonumber
\times
\frac{m_{e}e^4}{2 \omega _{1}\omega _{2}}|{\cal M}(k_1,k_2)|^2
\frac{d^3\textbf{k}_2}{(2\pi)^2}
.
\end{eqnarray}

We now specialize to a reference frame where the electron initially at rest, i.e. $ \textbf{p}_{i}=0 $ and $E_{i}=m_{e}$. Then the energy conservation law given by the argument of delta-function in Eq. (\ref{9}) is
\begin{eqnarray}
\label{10}
(p_i+k_1-k_2)^2-m_e^2
\\\nonumber
=2m_{e}(\omega_1-\omega_2)-2\omega_1\omega_2(1-\cos \theta_{12})=0
,
\end{eqnarray}
and the frequency of outgoing photon $ \omega_{2} $ can be expressed via $ \omega_{1} $ and the angle $ \theta_{12} $ between corresponding photon wave-vectors $ \textbf{k}_1$ and $\textbf{k}_2$ as follows
\begin{eqnarray}
\label{11}
\omega_2=\frac{\omega_1}{1+\frac{\omega_1}{m_{e}}(1-\cos\theta_{12})}
.
\end{eqnarray} 
Finally, performing summation over the spin and polarizations of final states and averaging over the spins and polarizations of initial states the differential cross section takes the form
\begin{eqnarray}
\label{12}
\frac{d\sigma_{\mathrm{sc}}}{d\Omega_2}
=
\frac{e^4}{16 \pi ^2}\left(\frac{\omega _{2}}{\omega_1}\right)^2
X(\omega_{1},\omega_{2})
,
\end{eqnarray}
together with notations
\begin{eqnarray}
\label{13}
X(\omega_{1},\omega_{2})=\frac{1}{2m_{e}^2}
F(\omega_1,\omega_2)
\\\nonumber
\times
\left(\Gamma ^4+16 m_{e}^2
   \left(\Gamma ^2+ 4\omega_1^2\right)+16 \Gamma ^2 m_{e} \omega_1\right)^{-1} 
\\\nonumber
\times   
   \left(\Gamma ^4+16 m_{e}^2 \left(\Gamma ^2+4\omega_2^2\right)-16 \Gamma ^2 m_{e} \omega_2\right)^{-1}
 ,
\end{eqnarray}
and
\begin{eqnarray}
\label{14}
F(\omega_1,\omega_2) = 
64 m_{e} \left(\Gamma ^6 (\omega_1-\omega_2)
\right.
\\\nonumber
+4\Gamma ^2 m_{e}^2 (\omega_1-\omega_2) \left(9 \Gamma ^2+4\omega_1^2+\omega_2^2\right)
\\\nonumber
+8 m_{e}^3 \left(4\Gamma ^2 \left(\omega_1^2-4 \omega_1
\omega_2+\omega_2^2\right)+5 \Gamma ^4
\right.
\\\nonumber
\left. +16\omega_1 \omega_2  \left(\omega_1^2+\omega_2^2\right)\right)
+128 m_{e}^5 \left(\Gamma ^2+(\omega_1-\omega_2)^2\right)
 \\\nonumber  
 +128 m_{e}^4 (\omega_1-\omega_2) \left(\Gamma ^2 -2\omega_1 \omega_2\right)
 \\\nonumber
 \left. 
   +2\Gamma ^4 m_{e} \left(\Gamma ^2+4\omega_1^2 -6 \omega_1 \omega_2 + 4  \omega_2^2\right)\right)
.   
\end{eqnarray}
In these equations the frequency $\omega_2$ of outgoing photon is given by Eq. (\ref{11}). 

It is easy to check that in the limit $ \Gamma\rightarrow 0 $ Eq. (\ref{12}) turns to the ordinary Klein-Nishina formula for monocromatic Compton scattering cross section
\begin{eqnarray}
\label{15}
\frac{d\sigma_{\mathrm{sc}}}{d\Omega_2}\equiv \lim\limits_{\Gamma\rightarrow 0}\frac{d\sigma_{\mathrm{sc}}^{\Gamma}}{d\Omega_2}
=\frac{r_{e}^2}{2}\left(\frac{\omega _{2}}{\omega_1}\right)^2
\\\nonumber
\times
\left(\frac{\omega_2}{\omega_1}+\frac{\omega_1}{\omega_2} - \sin^{2}\theta_{12}  \right)
,
\end{eqnarray}
where $ r_{e}=e^2/(4\pi m_{e})  $ is the classical electron radius. 

\begin{figure}[hbtp]
\centering
\includegraphics[scale=0.75]{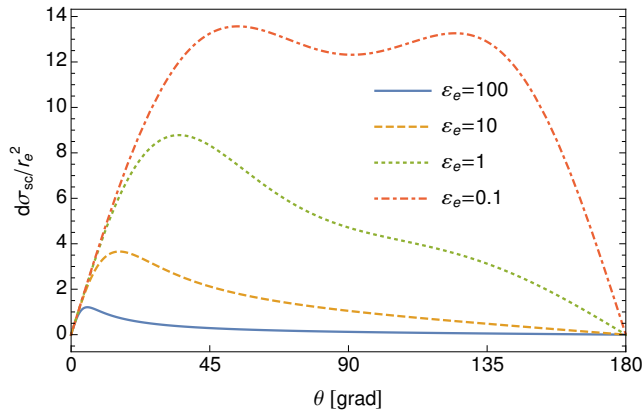}
\caption{The dependence of differential Compton scattering cross section on the angle $\theta_{12}$ between incident and outgoing photon (see Eq.~(\ref{15})). Here  $ r_{e}=e^2/(4\pi m_{e}) $ and the photon energy is measured in units of electron rest energy $\varepsilon_{e}\equiv \omega_{1}/m_{e}$.}
\label{fig1}
\end{figure}

\begin{figure}[hbtp]
\centering
\includegraphics[scale=0.75]{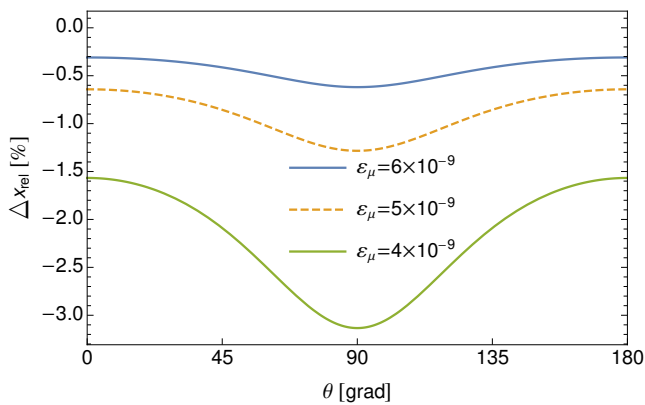}
\caption{Relative difference $\Delta \mathrm{x}_{\mathrm{rel}}$ in percent between modified and ordinary Compton cross sections for photon scattering on free muon (see Eqs. (\ref{12}) and (\ref{15}), respectively). Here $\varepsilon_{\mu}\equiv \omega_{1}/m_{\mu}$ and parameter $\Gamma$ in Eqs.~(\ref{13}), (\ref{14}) is $\Gamma_{\mu}=2.996\times 10^{-10}$ eV.}
\label{fig2}
\end{figure}

The angular dependence of ordinary Compton cross section Eq. (\ref{16}) for the different energies of incident photon is presented in Fig. \ref{fig1}. These results can be compared with the equation (\ref{12}) for finite parameter $\Gamma$. Within the considered approach which is based on the modified form of fermion propagator given by Eq. (\ref{1}), the scattering process of a quasi-monochromatic photon with a width $\Gamma$ on stable particle is equivalent to the scattering of a monocromatic photon by an unstable fermion with lifetime $\tau=1/\Gamma$.  As an example we consider the single Compton process for the muon with mass $m_{\mu}\approx 207m_{e}$ and decay time $\tau_{\mu} = 2.197\times 10^{-6}$ s \cite{MILOTTI1998137}. In Fig. \ref{fig2} the relative difference between ordinary and modified Compton cross sections is presented. 

It is important to consider the nonrelativistic limit of the above equations, which corresponds to the Thomson scattering. In this case $  \omega_1 / m_{e} \ll   1$ and the energy conservation law Eq. (\ref{11}) is $\omega_1\approx \omega_2$. Then setting $\omega_1= \omega_2$ in Eq. (\ref{15}) and performing the angular integration we arrive at \cite{weinberg1995quantum, Berest}
\begin{eqnarray}
\label{16}
\sigma_{\mathrm{Th}}=\frac{8\pi}{3}r_{e}^{2}
.
\end{eqnarray}

We will be interested in a similar expression but for a finite parameter $ \Gamma $. Keeping the terms of the oder $ \Gamma^{2} $ in the  Eq. (\ref{15}), integrating over the angles and taking the nonreltivistic limit we have
\begin{eqnarray}
\label{17}
\sigma_{\mathrm{Th}}^{\mathrm{\Gamma}} = \frac{\sigma_{\mathrm{Th}}}{1+\Sigma^{\Gamma}(\omega_1)}
,
\end{eqnarray}
where
\begin{eqnarray}
\label{18}
\Sigma^{\Gamma}(\omega_{1})=2 \pi  \Gamma ^2 
\left(70 m_{e}^2 \omega_{1}^2+120 m_{e}^3 \omega_{1}+60 m_{e}^4
\right.
\\\nonumber
\left.
+10 m_{e} \omega_{1}^3-13 \omega_{1}^4\right)(15 m_{e}^2 \omega_{1}^4)^{-1}
.
\end{eqnarray}

This result Eq. (\ref{17}) can be compared with one obtained from the classical electrodynamics with a radiation damping effects taking into account which is \cite{Heinzl, schwinger1998classical}
\begin{eqnarray}
\label{classic}
\sigma_{\mathrm{Th}} = \frac{\sigma_{\mathrm{Th}}}{1+(\gamma/\omega_{1})^2}
.
\end{eqnarray}
From comparison of Eqs. (\ref{17}) and (\ref{classic}) it follows that QED derivation leads to a different power-law dependence of Eq. (\ref{17}) on the incident radiation frequency $\omega_{1}$. It is important to note that the finite time effects in Thomson scattering have not been previously considered within the framework of the QED formalism. Existing approaches to solving this problem refer only to classical electrodynamics.

\section{Double compton scattering}
\label{double}

In this section, we will carry out similar calculations for the double Compton effect.
To begin with, we recall a brief derivation of the scattering cross section of the process $ \gamma(k)+e^{-}\rightarrow \gamma(k_1)+\gamma(k_2) + e^{-} $ \cite{mandl1952,JauchRohrlich}.
Following \cite{JauchRohrlich} the double Compton (DC) effect is described by the third order S-matrix element 
\begin{eqnarray}
\label{mandl1}
S^{(3)}_{fi}=(2\pi)^4\delta^{(4)}(p_{f}+k_1+k_2-p_{i}-k)
\\\nonumber
\times
\frac{m_{e}e^3}{\sqrt{8 V^5 E_{i}E_{f}\omega  \omega _{1}\omega _{2}}}{\cal M}(k,k_1,k_2)
,
\end{eqnarray}
where $ \cal{M} $ is the Feynman amplitude of the process 
\begin{widetext}
\begin{eqnarray}
\label{mandl2}
{\cal M}(k,k_1,k_2)=
\overline{u}(p_{f})
\left\lbrace
\hat{\varepsilon}(k)
S_{F}(k-p_{f})
\hat{\varepsilon}^*(k_{1})
S_{F}(k-k_1-p_{f})
\hat{\varepsilon}^*(k_{2})
+(1\leftrightarrow 2)
\right.
\\\nonumber
\left.
+
\hat{\varepsilon}^*(k_1)
S_{F}(-k_1-p_{f})
\hat{\varepsilon}(k)
S_{F}(k-k_1-p_{f})
\hat{\varepsilon}^*(k_{2})
+(1\leftrightarrow 2)
\right.
\\\nonumber
\left.
+
\hat{\varepsilon}^*(k_1)
S_{F}(-k_1-p_{f})
\hat{\varepsilon}^*(k_2)
S_{F}(-k_1-k_2-p_{f})
\hat{\varepsilon}(k)
+(1\leftrightarrow 2)
\right\rbrace
u(p_{i})
.
\end{eqnarray}
\end{widetext}
In the reference frame where the electron initially at rest, i.e. $ \textbf{p}_{i}=0 $ and $E_{i}=m_{e}$, the energy conservation law takes the form
\begin{eqnarray}
\label{mandl9}
\omega_2=\frac{m_{e} (\omega- \omega_{1})-\omega \omega_{1}(1-\cos \theta_1)}{m_{e}+\omega(1- \cos \theta_2)-\omega_{1}(1-\cos
   \theta_{12})}
.
\end{eqnarray}
After the summation over the spin and polarizations of final states, averaging over the spins and polarization of initial states and integrating over $\omega_{2}$ in Eq. (\ref{mandl1}) the differential cross section is
\begin{widetext}
\begin{eqnarray}
\label{mandl11}
\frac{d\sigma_{\mathrm{dc}}}{d\omega_1 d\Omega_1 d\Omega_2}
=
\frac{m_{e}e^6}{2^8\pi^5 }\left(\frac{\omega _{1}}{\omega}\right)\frac{m_{e} (\omega- \omega_{1})-\omega \omega_{1}(1-\cos \theta_1)}{(m_{e}+\omega(1- \cos \theta_2)-\omega_{1}(1-\cos
   \theta_{12}))^2}  
  X(\omega, \omega_1) 
,
\end{eqnarray}
\end{widetext}
where 
\begin{eqnarray}
\label{mandl12}
X(\omega, \omega_1) =\frac{1}{4}\mathrm{Tr}\sum\limits_{\mathrm{polarizations}}|{\cal M}(\omega, \omega_1)|^2 
.
\end{eqnarray}

\begin{figure}[hbtp]
\caption{Energy spectra of double Compton scattering cross section for given directions of both photons. The incident energy is $\varepsilon_{e}=\omega/m_{e}=2$ (1 MeV). The solid line corresponds to the angles $\theta_{1}=\theta_{2}=\frac{\pi}{2}$, $\theta_{12}=\frac{\pi}{2}$, while the dashed line to $\theta_{1}=\theta_{2}=\frac{\pi}{2}$, $\theta_{12}=0$.}
\centering
\includegraphics[scale=0.75]{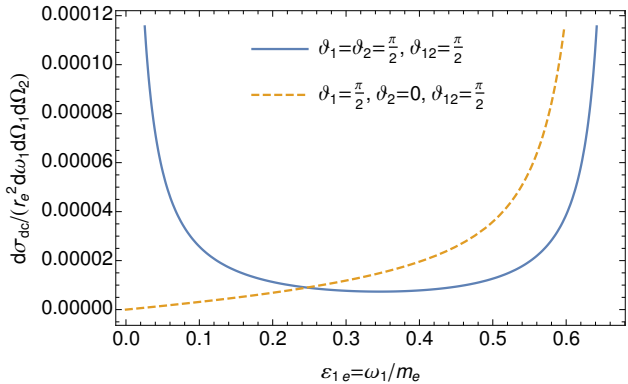}
\label{fig3}
\end{figure}

Summation in Eq. (\ref{mandl11}) and algebra with Dirac matrices can be performed in a fully analytical way with the use of {\it FeynCalc} software \cite{FeynCalc1, FeynCalc2}. To demonstrate the dependence of Eq. (\ref{mandl11}) on parameter $\Gamma$ we follow the examples considered in \cite{mandl1952, JauchRohrlich}. The results of numerical evaluation of double Compton scattering differential cross section for the energy $\varepsilon_{e}=\omega/m_{e}=2$ (1 MeV) of incident photon and $\Gamma=0$ are presented in Fig. \ref{fig3}. In this case the following infrared singularities occurs: two for the angles $\theta_{1}=\theta_{2}=\frac{\pi}{2}$, $\theta_{12}=\frac{\pi}{2}$ at $\varepsilon_{e}=0$ and $\varepsilon_{e}=2/3$, and for the angles $\theta_{1}=\frac{\pi}{2}$, $\theta_{2}=0$, $\theta_{12}=\frac{\pi}{2}$ at $\varepsilon_{e}=0$. 

To consistently cure the issue, the DC process has to be treated together with the next to leading order (NLO) radiative corrections to the Compton process \cite{feynman, chluba2}. In this context, one is usually interested in the total Compton scattering cross-section at order $\alpha^3$. NLO corrections display a logarithmic divergence which can be shown to cancel with the corresponding DC one. Employing a regularization - introducing a photon mass in the standard approach \cite{feynman}  - in the calculation of both cross-sections and summing the two, the dependence on the regularization parameters drops out, leaving a finite radiative correction at order  $\alpha^3$. The drawback of this approach is that the radiative correction cross-section now depends on the energy resolution of the experiment, $\omega_{\mathrm{res}}$. The argument to justify summing the two processes is that below some $\omega_{\mathrm{res}}\ll 1$ the experiment is unable to distinguish contributions from virtual photon emission, relevant to the computation of the NLO correction, and from real photon emission by DC. Adding all contributions to the total CS scattering cross-section it was found that the correction can exceed the naive $\alpha/\pi$ level at sufficiently high energies \cite{Mork}.

However, all of the regularization procedures are related to the consideration of the final state of the system \cite{feynman}. Alternative physical approach for solving this problem based on the analysis of the input photon - taking into account its finiteness in time. Indeed, the standard consideration assumes that the incident photon is taken in the form of a monochromatic plane wave infinite in time and space \cite{Akhiezer, bjorken1964relativistic, Landau, Berest}. Only in this case there is possibility to form an arbitrarily low-frequency photon at the end of the process. However, in reality, there are no such waves in nature. There are only real photons, which are formed by some radiation mechanism during a finite time interval \cite{dawson1970}. As a result, the scattering of such a photon by an electron lead to a finite spectrum that is completely finite in the entire frequency range. Recently in \cite{physics3040074} it was shown that the scattering cross section for inverse double Compton (IDC) process have no ordinary infrared divergence related to the resolution of photon detector in final states. Instead of this the singularity is moved to the incident photon energies $\omega_{1}$ and $\omega_{2}$  which are the input parameters for differential IDC cross section. From consideration of the inverted process, it becomes obvious that restrictions must be imposed not on the resolution of the detector $\omega_{\mathrm{res}}$ (which hypothetically can be made arbitrarily sensitive to the low-energy part of the DC spectrum) but on the photon source. The latter means that the incident photon should actually be considered as non-monochromatic one.

\begin{figure}[hbtp]
\caption{Energy spectra of double Compton scattering cross section for given directions of both photons and different values of width parameter in units of electron mass $y=\Gamma/m_{e}$. The incident energy is $\varepsilon_{e}=\omega/m_{e}=2$ (1 MeV) and the angles are $\theta_{1}=\theta_{2}=\frac{\pi}{2}$, $\theta_{12}=0$. The solid blue line goes to infinity at $\varepsilon_{1e}=2/3$ (or $\varepsilon_{2e}=0$ according to energy conservation law Eq. (\ref{11})).}
\centering
\includegraphics[scale=0.75]{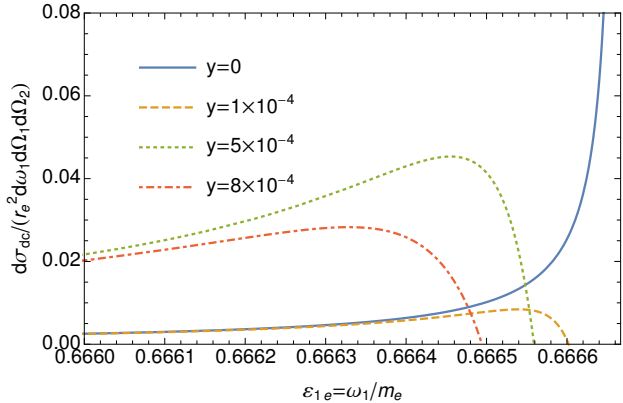}
\label{fig4}
\end{figure}

It is useful to compare the behaviour of DC differential cross section near the singularity for finite parameter $\Gamma\neq 0$ and limiting case $\Gamma=0$. The corresponding results are presented in Fig. \ref{fig4} for $\varepsilon_{e}=2$ (1 MeV) and angles $\theta_{1}=\theta_{2}=\frac{\pi}{2}$, $\theta_{12}=0$. From Fig. \ref{fig4} it is seen that in the case $\Gamma\neq 0$ the infinite peak arising as a result of infrared divergence is greatly smoothed out. 

\section{Results and discussion}
\label{conclusion}

In this work, we analyzed finite-time effects in scattering one or two photons on a free fermion particle. Localized in time interaction was introduced by a small complex addition to the photon momentum (which means going beyond the monochromatic plane-wave approximation \cite{Zalialiutdinov_2014}). To do this, we used the theory of propagation functions of unstable fermions, which also contain a complex addition to the momentum related to the particle's lifetime \cite{weinberg1995quantum, Kuksa2015}. At sufficiently low incident photon energies but still much larger than the corresponding pulse width, the cross sections of the considered processes can differ significantly compared to the standard approach. The expressions obtained in this work can also be applied to the scattering of photons by unstable particles, where the particle's lifetime plays the role of inverse width. The considered approach also leads to the regularization of the infrared divergence in the cross section of the double Compton effect in complete analogy with bound-state QED, where the natural width of atomic level plays the role of regularising parameter.

It is worth noting that other approaches for introducing a finite interaction time have also been discussed in a recent work \cite{Correlated}. In particular, in \cite{Correlated} it was proposed to multiply the amplitude of the process, Eq. (\ref{mandl2}), by a regularizing factor containing the parameter of a pulse duration from the source.

Considering the above, it becomes evident that with an increase in the detectors sensitivity to the low energy part of the spectrum, the cross section of double Compton effect in the soft photon region can be studied experimentally. In particular, direct observation of peak flattening from infrared divergence, see Fig. \ref{fig4}, could verify the proposed theoretical approach. An experimental study of the considered effects can be carried out with optical photons. For the double Compton effect, this will result in the reemitted soft photon being in the radio-frequency range, where sufficiently sensitive detectors are available.

\section{Acknowledgements}
The authors are grateful to Dr. V. Zalipaev (Laboratory of Quantum Processes and Measurements, ITMO University, St. Petersburg, Russia) and Dr. D. Solovyev (Department of Physics, St. Petersburg State University, Russia) for valuable discussion.

\bibliography{mybibfile} 
\end{document}